\documentclass[twocolumn,american,english,aps,prb,showpacs,showkeys,amsmath,amssymb,superscriptaddress,a4]{revtex4-1}
\usepackage[T1]{fontenc}
\usepackage[latin9]{inputenc}
\usepackage{color}
\usepackage{array}
\usepackage{booktabs}
\usepackage{textcomp}
\usepackage{multirow}
\usepackage{amstext}
\usepackage{amssymb}
\usepackage{graphicx}
\usepackage{esint}
\usepackage{subscript}

\makeatletter

\providecommand{\tabularnewline}{\\}

\@ifundefined{textcolor}{}
{%
 \definecolor{BLACK}{gray}{0}
 \definecolor{WHITE}{gray}{1}
 \definecolor{RED}{rgb}{1,0,0}
 \definecolor{GREEN}{rgb}{0,1,0}
 \definecolor{BLUE}{rgb}{0,0,1}
 \definecolor{CYAN}{cmyk}{1,0,0,0}
 \definecolor{MAGENTA}{cmyk}{0,1,0,0}
 \definecolor{YELLOW}{cmyk}{0,0,1,0}
}

\usepackage{multirow}\usepackage{bm}\usepackage{gensymb}\usepackage{threeparttable}

\makeatother

\usepackage{babel}
\begin{document}
\selectlanguage{american}%

\title{The origin of the vanadium dioxide transition entropy}

\author{Thomas A. Mellan}
\email{t.mellan@imperial.ac.uk}

\affiliation{Thomas Young Centre for Theory and Simulation of Materials, Department
of Materials, Imperial College London, Exhibition Road, London SW7
2AZ, United Kingdom}

\author{Hao Wang}
\author{Udo Schwingenschl\"ogl}
\affiliation{King Abdullah University of Science and Technology (KAUST),
Physical Science and Engineering Divison (PSE), Thuwal 23955-6900,
Saudi Arabia}

\author{Ricardo Grau-Crespo}

\affiliation{Department of Chemistry, University of Reading, Whiteknights, Reading RG6 6AD, United Kingdom}

\date{\today}

\begin{abstract}

The reversible metal-insulator transition in VO\textsubscript{2} at $T_\text{C} \approx 340$ K has been closely scrutinized yet its thermodynamic origin remains ambiguous. We discuss the origin of the transition entropy by calculating the electron and phonon contributions at $T_\text{C}$ using density functional theory.  The vibration frequencies are obtained from harmonic phonon calculations, with the soft modes that are imaginary at zero temperature renormalized to real values at $T_\text{C}$ using experimental information from diffuse x-ray scattering at high-symmetry wavevectors. Gaussian Process Regression is used to infer the transformed frequencies for wavevectors across the whole Brillouin zone, and in turn compute the finite
temperature phonon partition function to predict transition thermodynamics. Using
this method, we predict the phase transition in VO\textsubscript{2}
is driven five to one by phonon entropy over electronic entropy, and predict a total transition entropy that accounts for $95 $\% of the calorimetric value.

\end{abstract}
\selectlanguage{american}%

\keywords{metal-insulator transition, thermodynamics, phonon entropy, VO\textsubscript{2},
Gaussian Process Regression}

\maketitle
\selectlanguage{american}%
\begin{figure*}
\begin{centering}
\includegraphics[scale=0.9]{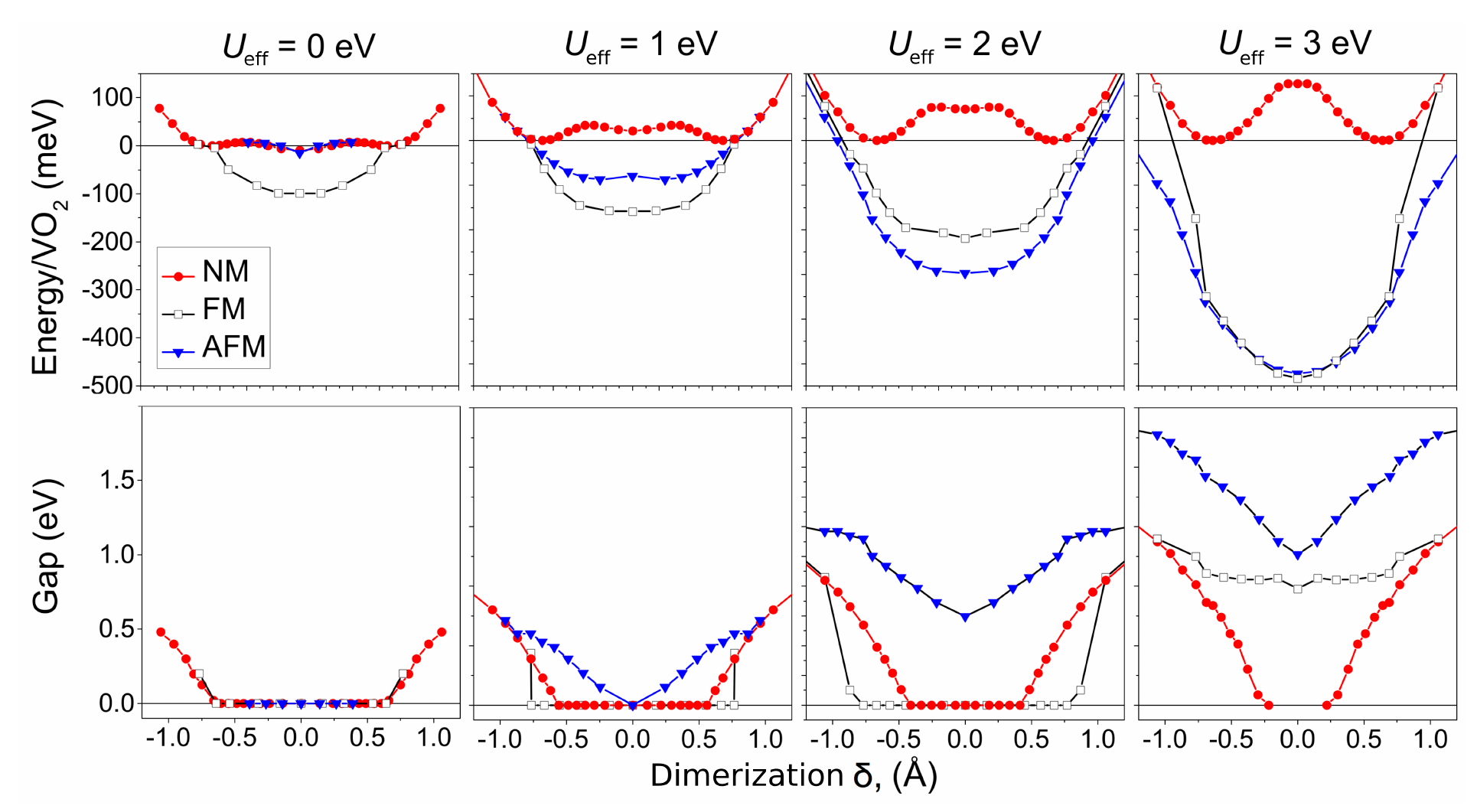}
\par\end{centering}
\caption{{Potential energy surface (eV/VO\protect\textsubscript{2}) and band
gap (eV) as a function of dimerization $\delta\,(\text{Å})$, which is the difference in $d(\text{V-V})$ between consecutive pairs of cations along the rutile c axis. Values
are shown for non-magnetic (NM), antiferromagnetic (AFM) and ferromagnetic
(FM) ordering, and for a range of \emph{d }electron on-site Coulomb
interaction strengths, $U_{\text{eff}}=[0,3]$ eV. The NM $U_{\text{eff}}=3$ eV
description appropriately opens the band gap in dimerization and provides
a mechanically unstable high-symmetry phase. \label{fig:Full PES}}}
\end{figure*}

The first-order phase transition in VO\textsubscript{2} occurs at a temperature of 
$T_{\text{C}}\approx 340\,\text{K}$, and is coupled to defect concentration,\cite{Manning2004a,Netsianda2008,Tan2012,Piccirillo2008,Jin1998,Warwick2014}
strain field,\cite{Tselev2010,Merced2013,Transition1970}, electric field and optical
fluence.\cite{OCallahan2015a,Chudnovskiy2002,Cavalleri2001} The transition has
been studied since Klemm and Grimm in the 1930's,\cite{Klemm1939}
Cook in the 40's,\cite{Cook1947a} and in detail by Morin in 1959.\cite{Morin1959}
Fundamental questions on the nature of the transition have been debated
for decades,\cite{E1975,Paquet1980,Eguchi2008,Wentzcovitch1994} and
continue to be researched.\cite{Zheng2015,Biermann2005,Wall2018} The transition
occurs most notedly in temperature so understanding the thermodynamic
origin is a point of basic importance. 

In this study we use density functional theory (DFT) to predict the origin of the VO\textsubscript{2} transition entropy. The applicability
of DFT to describe the transition metal oxide class of solids depends
sensitively on technical details.\cite{grau2013examining,Xiao2014}
We use non-spin-polarized calculations based on the PBE exchange correlation functional,\cite{Perdew1996} with on-site Coulomb
correction $U_\text{eff}=3$ eV.\cite{Dudarev1998}  As shown in Fig. \ref{fig:Full PES}, this approach leads to agreement with experiment on the following important points:
\begin{enumerate}
\item Electronic structure -- the high-symmetry metallic R phase is appropriately
gapless. Band gap is opened smoothly with V-V dimerization, resulting
in a semiconducting monoclinic (M1) phase.
\item Transition enthalpy -- the low-temperature M1 phase is energetically favored
over the high-temperature R phase.\cite{energyNote}
\item Mechanical stability -- the low-symmetry M1 phase is stable against
distortion and the high-symmetry R phase is unstable in $0$ K DFT
simulation.
\end{enumerate}
Including spin polarization is shown in Fig. \ref{fig:Full PES}  to lower the DFT energy of R-VO\textsubscript{2} with respect to the non-magnetic solution, destroying agreement with experiment for the points listed above. The problems related with spin polarization in the DFT description of VO\textsubscript{2} have been discussed before,\cite{Grau-Crespo2012} and have been resolved fully only in the context of Quantum Monte Carlo simulations,\cite{Zheng2015} which are too computationally expensive to use to investigate lattice dynamics. We therefore take the pragmatic approach employed by other authors of using non-magnetic calculations,\cite{Eyert2011} on the basis on agreement with experiment.

For the M1 phase, the Born-Oppenheimer surface is convex about equilibrium
coordinates. The harmonic approximation to the interatomic potential
is appropriate for small displacements, and is expected to be adequate
for M1-VO\textsubscript{2} up to $T_{\text{C}}$. On the other hand the high-symmetry
R phase has negative second-order force constants which qualitatively
invalidate free energy predictions at the harmonic level.
Approaches to remedy this that include anharmonic effects have become more accessible thanks to recent developments,\cite{Duff2015,Hickel2011,Hellman2013,Zhou2014,Prentice2017,Monserrat2013} enabling the description of systems with light atoms, at ultra-high temperatures, or near phase
transitions, yet widespread application of first principles anharmonic thermodynamics
remains limited due to computational cost and complexity.
In this work we present a simple, experimentally-motivated approach to compute the thermodynamics of temperature-stabilised imaginary modes
in VO\textsubscript{2}. The method is low-cost and applicable generally
to the DFT thermodynamics of high-temperature phases that are unstable
at zero temperature.

\begin{figure*}
\centering{}\includegraphics[scale=0.75]{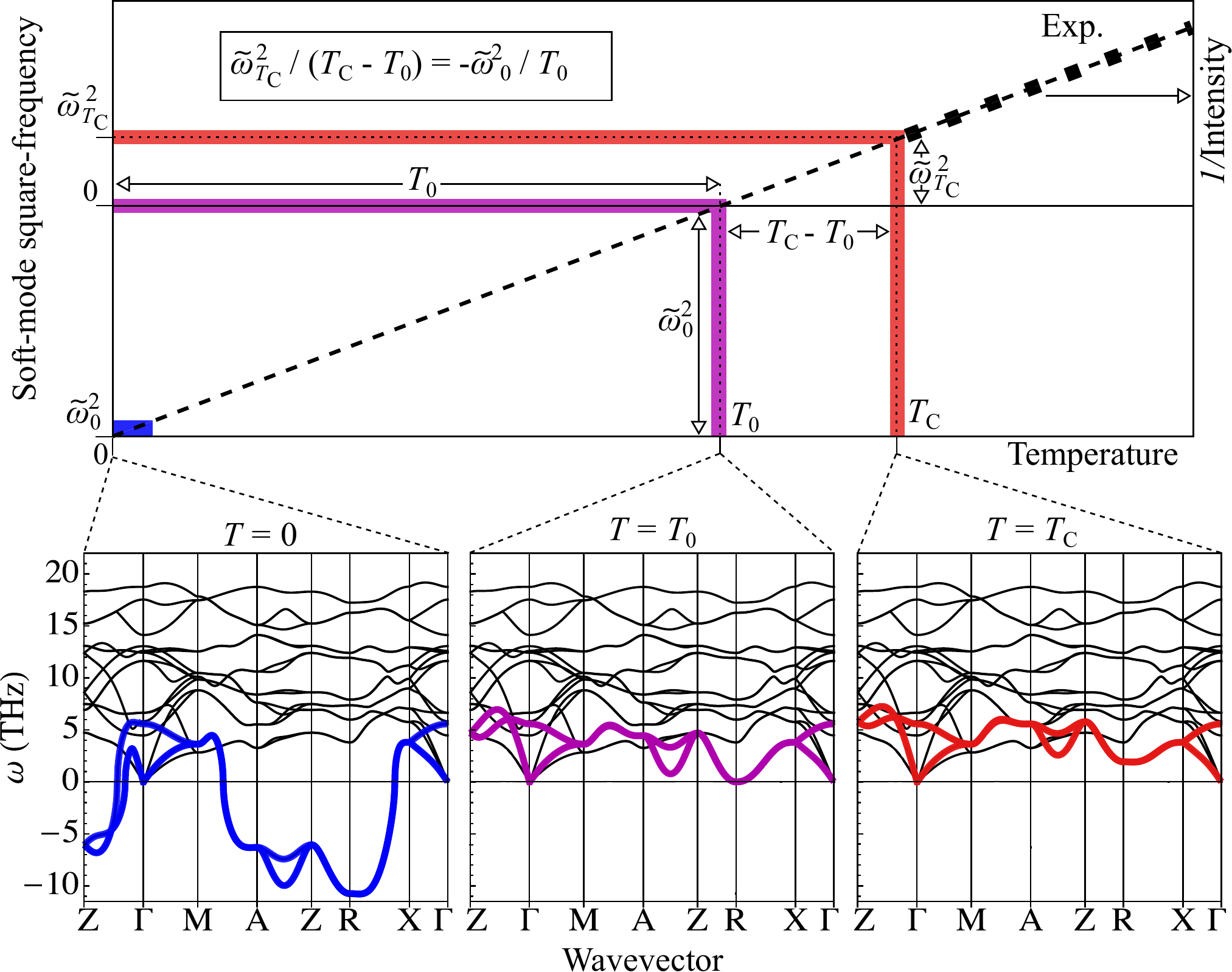}\caption{{\emph{Top,} \emph{right axis}: Inverse intensity from diffuse x-ray
scattering versus temperature{.\cite{Terauchi1978}}
\emph{Top,} \emph{left axis:} Linear softening of square frequencies
with temperature. The transition occurs at $T_{\text{C}}$ and
the classical second-order transition temperature is $T_{0}$. The
frequencies at $0$ K and $T_{\text{C}}$, shown as $\tilde{\omega}_{0}$
and $\tilde{\omega}_{T_{\text{C}}}$, are related linearly. \label{fig: SoftModeTemperatureDependence}
\emph{Bottom:} DFT phonon dispersion for R-VO\protect\textsubscript{2}
at 0 K, $T_{0}$ and $T_{\text{C}}$. Thick colored lines show the
soft modes after renormalization, at 0 K (blue), $T_{0}$ (magenta)
and $T_{\text{C}}$ (red).}}
\end{figure*}

In the soft mode theory of Cochran and in Landau phenomenological
approaches,\cite{Cochran1959,Cochran1961,Cochran1981,Landau1980}
a square-root temperature dependence is identified for transition
parameters. The squared-frequency $\tilde{\omega}_{i\mathbf{q}}^{2}$ of a mode $i$ that softens near the transition at wavevector $\mathbf{q}$ is expected to  decrease linearly with temperature towards a first-order solid-state transition:\cite{CochranFRS1973}

\begin{equation}
\tilde{\omega}_{i\mathbf{q}}^{2}(T)\propto T-T_{0}\,.\label{eq: linearResponse-1}
\end{equation}
In VO\textsubscript{2} this proportionality has been observed in
experimental measurements.\cite{Terauchi1978} For example, Cohen and Terauchi report a linear temperature
response from diffuse x-ray scattering measurements at $\mathbf{q}=\mathbf{R}$ with $\mathbf{R}=\left\langle \frac{1}{2}0\frac{1}{2}\right\rangle $ which is represented in the left axis in Fig. \ref{fig: SoftModeTemperatureDependence}.
The temperature $T_0$, when $\tilde{\omega}_{i\mathbf{R}}\to0$,  has a value of $T_{0}=329\,\text{K}$, and corresponds to the classical second-order transition temperature.
Along with the first-order transition temperature $T_{\text{C}}$, Eqn. \ref{eq: linearResponse-1} relates the phonon frequency at 0 K, $\tilde{\omega}_{i\text{\ensuremath{\mathbf{R}}}}^{2}(0)$, to the frequency at the transition temperature as
\begin{equation}
\tilde{\omega}_{i\text{\ensuremath{\mathbf{R}}}}^{2}(T_{\text{C}})=\tilde{\omega}_{i\text{\ensuremath{\mathbf{R}}}}^{2}(0)\frac{T_{\text{C}}-T_{0}}{-T_{0}}\,.\label{eq: omegaAtTc}
\end{equation}
Here the shifted frequency $\tilde{\omega}_{i\text{\ensuremath{\mathbf{R}}}}(0)$ is equal to the harmonic frequency $\omega_{i\mathbf{R}}$ that is calculated with DFT.
The application of the transformation of imaginary harmonic DFT
frequencies at $0$ K to real frequencies at $T_{\text{C}}$ is shown in Fig. \ref{fig: SoftModeTemperatureDependence}. The method gives values for the temperature-stabilised frequencies at negligible additional cost to standard harmonic DFT calculations, provided the coefficients $T_{0}$
and $T_{\text{C}}$ are known,  which is commonly the case as shown by the experimental data reviewed by Cochran and Cowley.\cite{CochranFRS1973,Landau1980}  

To implement the frequency shifts for R-VO$_2$, the transition modes which are shown in  Fig. \ref{fig: SoftModeTemperatureDependence} to be imaginary at $\mathbf{R},$ $\mathbf{Z}$ and $\mathbf{A}$ in $q_{z}=\frac{1}{2}$, are renormalized to $T_\text{C}$ following the prescription described in the previous paragraph. At other R-VO$_2$ wavevetors (see Brillouin zone geometry, Fig. \ref{fig:M1 R phonon}, Appendix), the frequencies do not soften to imaginary harmonic frequencies. For example, in Fig. \ref{fig: SoftModeTemperatureDependence} the $\mathbf{\Gamma}$,  $\mathbf{X}$,
and $\mathbf{M}$ wavevectors in $q_{z}=0$  have real harmonic  frequencies at $T=0$  K. The frequencies  in  $q_{z}=0$ that do not soften are modelled using the DFT harmonic frequencies $\omega_{i\mathbf{q}}$.

In order to make thermodynamic predictions for a high-temperature
phase we need to sample the transformed modes finely across the Brillouin zone, not only at the limited high-symmetry $\mathbf{q}$-points in the $q_{z}=\frac{1}{2}$ and $q_{z}=0$ regions described. 
To obtain $\text{\ensuremath{\tilde{\omega}_{i\mathbf{q}}}}(T_{\text{C}})\in\mathbb{R\,}\forall\,i\mathbf{q}$, the partial knowledge we already have of the frequencies transformed to $T_\text{C}$ is found to be sufficient data for machine learning
techniques to interpolate $\tilde{\omega}_{i\mathbf{q}}(T_{\text{C}})$ to
arbitrary phonon wavevectors. $\tilde{\omega}_{i\mathbf{q}}(T_{\text{C}})$ is
inferred at all irreducible Brillouin zone wavevectors using Gaussian
Process Regression (GPR),\cite{rasmussen2006gaussian} enabling the partition
function of the R-VO\textsubscript{2} vibrational system to be specified
at $T_{\text{C}}$. GPR accuracy benchmarks and technical details
are provided in Appendix \ref{sec: GPR}.

To understand the source of entropy driving the transition, we compute
$S_{\text{DFT}}=S_{\text{R}}-S_{\text{M1}}$ at $T_{\text{C}}$. $S_{\text{M1}}$
is the DFT harmonic vibrational entropy of M1-VO\textsubscript{2},
and for R-VO\textsubscript{2} the entropy is calculated as $S_{\text{R}}=S_{\text{R}}^{\text{el}}+S_{\text{R}}^{\text{ph}}+\tilde{S}_{\text{R}}^{\text{ph}}$
with consecutive terms from electrons, the harmonic phonon entropy,
and the soft-mode phonon entropy from the two experimentally-renormalized
transition modes. Thermodynamic calculation details are provided in
Appendix \ref{sec:thermodynamics}.

The total transition entropy we predict for VO\textsubscript{2} is
$S_{\text{DFT}}=1.42$ \emph{k}\textsubscript{B}/VO\textsubscript{2}.
The commonly referenced calorimetric value is $S_{\text{exp}}=1.5$
\emph{k}\textsubscript{B}/VO\textsubscript{2}.\cite{Ber1969} Our
predicted value of $S_{\text{DFT}}=1.42$ \emph{k}\textsubscript{B}/VO\textsubscript{2} accounts for 95 \% of the calorimetric value. The predicted value
is composed of the contributions $S^{\text{el}}=0.25$ and $S^{\text{ph}}=1.17$
\emph{k}\textsubscript{B}/VO\textsubscript{2}. The source of entropy
driving the transition is therefore phonons over electrons at a ratio
of almost five to one.

In Table \ref{tab: Historical-context_VO2_entropies} our entropy predictions are compared to 13 historically reported values from the literature.
The values range widely, from $S^{\text{el}}=0.01$ to $0.6$ \emph{k}\textsubscript{B}/VO\textsubscript{2}
and $S^{\text{ph}}=0.64$ to $1.35$ \emph{k}\textsubscript{B}/VO\textsubscript{2}.       Among the range of predictions, our conclusions align most closely with those of Budai \emph{et al.},\cite{Budai2014a} with which we agree that the transition thermodynamics are mostly due to phonons. One difference is that Budai \emph{et al.}\cite{Budai2014a} compute a phonon entropy of $0.93$ \emph{k}\textsubscript{B}/VO\textsubscript{2} compared to $1.17$ \emph{k}\textsubscript{B}/VO\textsubscript{2} here. Their value is based on $U_\text{eff}=0$ eV calculations, whereas this work uses  $U_\text{eff}=3$ eV  DFT, to ensure a qualitatively correct description of the electron band gap, transition enthalpy, and R-point lattice instability. (Note, the sensitivity of $S^{\text{ph}}$  is less than 1 \%  per 0.1 eV of $U_\text{eff}$ about the appropriate value of $U_\text{eff}=3$  eV, but the $U_\text{eff}=0$  eV description has qualitatively incorrect features.) A second difference with the results of  Budai \emph{et al.}\cite{Budai2014a} concerns their reported 'best' predictions, which use a phonon entropy estimate from scattering measurements, which is $1.02\pm0.09$ \emph{k}\textsubscript{B}/VO\textsubscript{2}, compared to our value of $1.17$ \emph{k}\textsubscript{B}/VO\textsubscript{2}. Consequently the error with respect to the total calorimetric value\cite{Ber1969} of $1.5 \pm 0.01$  \emph{k}\textsubscript{B}/VO\textsubscript{2} is $0.08$ \emph{k}\textsubscript{B}/VO\textsubscript{2} here, rather than $0.21$ \emph{k}\textsubscript{B}/VO\textsubscript{2}.\cite{Budai2014a}

A controversial point that merits discussion is the possibility of
a spin contribution to the transition entropy. Quantum Monte Carlo calculations have
predicted that the R phase, which in nature only exists above $T_{\text{C}}$, would be spin ordered at $T=0$ K.\cite{Zheng2015} On this basis
Xia and Chen suggest a spin contribution to the transition entropy.\cite{Xia2017a}
Accounting for a coincident spin disordering at $T_{\text{C}}$ in
our predictions increases the entropy value by $\text{ln}\left(2\right)=0.69$
\emph{k}\textsubscript{B}/VO\textsubscript{2} to $S_{\text{DFT}}=S^{\text{el}}+S^{\text{ph}}+S^{\text{spin}}=0.25+1.17+0.69=2.11$
\emph{k}\textsubscript{B}/VO\textsubscript{2}, which exceeds the experimental
value of $S_{\text{exp}}=1.5$ \emph{k}\textsubscript{B}/VO\textsubscript{2}.
If a $\text{ln}\left(2\right)$ spin contribution to the transition
entropy exists, $S^{\text{ph}}$ must be considerably lower for $S_{\text{DFT}}$
to remain consistent with $S_{\text{exp}}$. Considering the neutron
scattering measurements by Budai \emph{et al}. we are inclined to believe this is
unlikely.\cite{Budai2014a} The neutron measured phonon density of states (DOS) can be
used to estimate a phonon entropy of $1.02$ \emph{k}\textsubscript{B}/VO\textsubscript{2}
(Table \ref{tab: Historical-context_VO2_entropies}), which is similar to our predicted value of $S^{\text{ph}}=1.17$ \emph{k}\textsubscript{B}/VO\textsubscript{2},
and insufficiently small to accommodate the full spin term. We therefore consider
 that a fully disordered Heisenberg spin contribution to the transition
unlikely. In order to confirm or refute inferences based on our experimentally-renormalized
DFT thermodynamics and the neutron scattering measurements of Budai
\emph{et al}.\cite{Budai2014a}, we propose a simple experiment to measure $T_{\text{C}}$ in the presence of a strong magnetic field. If there is a spin contribution to the entropy, it should vanish in the presence of the magnetic field, which will bring the value of $T_{\text{C}}$ significantly up. If there is no magnetic entropy involved in the transition, $T_{\text{C}}$ should not change or change very little in the presence of the field.

\begin{table*}
\caption{{Historical measured and computed VO\protect\textsubscript{2} transition
entropies, $S$ ($k_{\text{B}}$/VO\protect\textsubscript{%
2%
}), along with available partial electron, phonon and spin contributions.
$^{*}$Unpublished measurements by Ryder, reported by Berglund \emph{et
al.}.\cite{Ber1969} $^{\text{§}}$Values determined from analysis of Ryder's
measurements.\cite{Ber1969} {$^{\dagger}$}Mott
and Zylbersztejn base their analysis on a total transition entropy
of $S=1.6\,$ $k_{\text{B}}$/VO\protect\textsubscript{%
2%
}, mis-citing a Berglund report which has the entropy at $S=1020\pm5$
cal/mol or $S=1.51\pm0.01$ $k_{\text{B}}$/VO\protect\textsubscript{%
2%
}, assuming $T=340.5\pm0.5$ K. \label{tab: Historical-context_VO2_entropies} }}

\begin{centering}
\begin{tabular}{cccccc}
\toprule 
\multirow{2}{*}{Source} & \multirow{2}{*}{Method} & \multicolumn{4}{c}{Entropy contributions}\tabularnewline
 &  & $S^{\text{ph}}$  & $S^{\text{el}}$ & $S^{\text{spin}}$ & $S$\tabularnewline
\midrule
\midrule 
Klemm and Grimm,\cite{Klemm1939} 1939 & Calorimetry measurements & - & - & - & 1.2\tabularnewline
Cook\cite{Cook1947a}, 1947 & Calorimetry measurements & - & - & - & 1.50\tabularnewline
Kawakubo\cite{Kawakubo1965}, 1964 & Calorimetry measurements & - & - & - & 1.1\tabularnewline
Ryder\cite{Ber1969}, 1969 & Calorimetry measurements & - & - & - & $1.51\pm0.01$$^{*}$\tabularnewline
Berglund \emph{et al.}\cite{Ber1969}, 1969 & Analysis of Ryder's heat capacity measurements & 1.25$^{\text{§}}$ & 0.25$^{\text{§}}$ & - & $1.51\pm0.01$$^{*}$\tabularnewline
Paul\cite{Paul1970}, 1970 & Parabolic band model calc. & - & 0.15 & - & -\tabularnewline
Hearn\cite{Searcha}, 1972 & 1D model calc. & 1.17 & 0.01 & - & 1.18\tabularnewline
Chandrasekhar \emph{et al.}\cite{Chandrasekhar1973}, 1973 & Scanning calorimetry measurements & - & - & - & 1.65\tabularnewline
Zylbersztejn and Mott\cite{E1975}, 1975 & Analysis of magnetic susceptibility measurements& 1.02 & 0.58 & - & 1.6 (1.51)$^{\dagger}$\tabularnewline
Pintchovski \emph{et al.}\cite{Pintchovski1978}, 1978 & Calorimetry and electrical resistivity measurements & 0.9 & 0.6 & - & -\tabularnewline
Maurer \emph{et al.}\cite{Maurer1999}, 1999 & Debye model fitted to sound velocity measurement & 1.35 & - & - & -\tabularnewline
Budai \emph{et al}.\cite{Budai2014a}, 2014 & IXS phonon measurements and DFT electron calc.& $1.02\pm0.09$ & 0.27 & - & $1.29\pm0.09$ \tabularnewline
Budai \emph{et al}.\cite{Budai2014a}, 2014 &  DFT ($U=0$ eV) MD and DFT electron calc.& $0.93$ & 0.27 & - & $1.2$ \tabularnewline
Xia and Chen\cite{Xia2017a}, 2017 & Compressed sensing DFT  phonon and electron calc. & 0.64 & 0.25 & 0.69 & 1.58\tabularnewline
This work & Exp.-renormalized DFT phonon and electrons calc. & 1.17 & 0.25 & - & 1.42\tabularnewline
\bottomrule
\end{tabular}
\par\end{centering}
\centering{}
\end{table*}

\section*{Conclusion}

We have described the source of entropy driving the VO\textsubscript{2}
metal-insulator transition. Our thermodynamic predictions suggest the transition
is driven by phonons over electrons at a ratio of $1.17:0.25$, and that the computed entropy accounts for 95 \% of the calorimetric entropy value. In order to make our predictions we have performed DFT harmonic phonon calculations, in conjunction with an experimentally-motivated
soft-mode renormalization scheme based on data from x-ray scattering measurements.
The scheme has predicted values of soft-mode frequencies at the transition
temperature for high-symmetry points in the Brillouin zone of R-VO\textsubscript{2}.
The machine learning interpolation method Gaussian Process Regression
was used to infer the soft-mode frequencies across the full Brillouin
zone based on the input of frequencies at partial high-symmetry wavevectors.
A simple procedure has been proposed to experimentally confirm or
refute claims of a spin disorder contribution to the transition entropy.

\selectlanguage{american}%
\begin{acknowledgments}
R.G.C. and T.A.M. acknowledge funding from the UK\textquoteright s
Engineering and Physical Sciences Research Council EPSRC (EP/J001775/1).
Via the UK\textquoteright s HPC Materials Chemistry Consortium, which
is funded by EPSRC (EP/L000202), this work made use of ARCHER, the
UK\textquoteright s national high-performance computing services.
The research reported in this publication was supported by funding
from King Abdullah University of Science and Technology (KAUST).
T.A.M  is grateful for computational support from the UK Materials and
Molecular Modelling Hub, which is partially funded by EPSRC (EP/P020194),
for which access was obtained via the UKCP consortium and funded by
EPSRC grant ref EP/P022561/1.
\end{acknowledgments}

\clearpage{}

\rule[0.5ex]{1\columnwidth}{1pt}

\section*{References}

\bibliographystyle{unsrt}
\bibliography{VO2_PES_2018}

\begin{thebibliography}{10}
\bibitem{energyNote}
{The on-site Coulomb interaction could be tuned arbitrarily to reproduce the
  experimental latent heat of transition of 44 meV/VO$_2$, but such tuning is
  beyond the scope of this paper.}
\end{thebibliography}

\clearpage{}

\part*{Appendix\label{part:Supplementary-information}}

\section{DFT calculations\label{sec:DFT-calculations}}

Periodic DFT calculations were performed with the Vienna \textit{Ab-initio}
Simulation Package (VASP),\cite{Kresse1996a,Kresse1996} using the
generalized gradient approximation (GGA) in the form of the Perdew-Burke-Ernzerhof
exchange-correlation functional (PBE).\cite{Perdew1996} The projected
augmented wave method was used to describe the interaction between
the valence electrons and the core states, which were kept frozen
at the atomic references (up to $3$\emph{p} in V and $1$\emph{s}
in O).\cite{Blochl1994,Kresse1999} Plane waves were cutoff at a
kinetic energy of 520 eV, and $\mathbf{k}$-points were sampled at
a density of $6\times6\times9$ divisions per rutile unit cell. Force and energy
convergence thresholds were set to $10^{-3}$ eV/Å and $10^{-6}$
eV respectively.

The Coulomb interaction between vanadium \emph{d} electrons was corrected
with an effective on-site term, $U_{\text{eff}}$.\cite{Dudarev1998}
The effect of $U_{\text{eff}}$ and magnetic ordering was considered
for phase enthalpy and band gap.
NM $U_{\text{eff}}=3$ eV calculations reproduce the basic characteristics well
known from experiment, including instability of the high-symmetry
$P4_{2}/mnm$ rutile phase at low temperature, and electronic band
gap  phase opening with V-V dimerization.

\section{Phonon calculations\label{sec:Phonon-calculations}}

\begin{figure*}[p]
\begin{centering}
\includegraphics[scale=0.3]{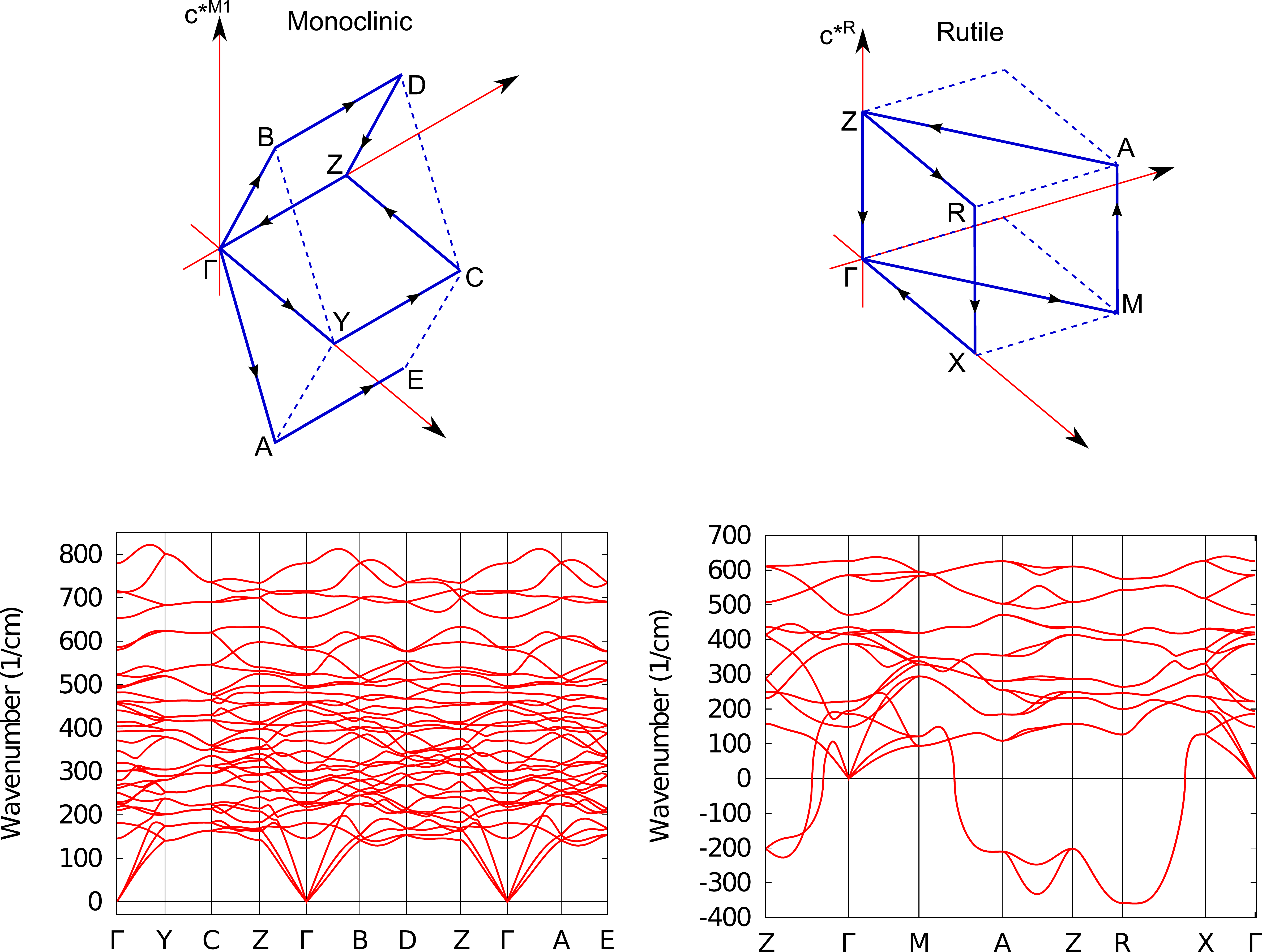}
\par\end{centering}
\caption{{DFT harmonic phonon dispersion for M1 and R-VO\protect\textsubscript{2},
and Brillouin zone sampling paths. \label{fig:M1 R phonon}}}
\end{figure*}

Phonons were computed from second-order force constants using the
\textsc{\footnotesize{}PHONOPY} code.\cite{Chaput2011} The M1 and
R phases employ $2\times2\times2$ and $2\times2\times3$ supercells respectively. Phonon thermodynamics
functions were satisfactorily converged at a sampling density equivalent
to $16\times16\times24$ $\mathbf{q}$-point mesh for the rutile conventional
unit cell.

\begin{figure}
\begin{centering}
\includegraphics[scale=0.66]{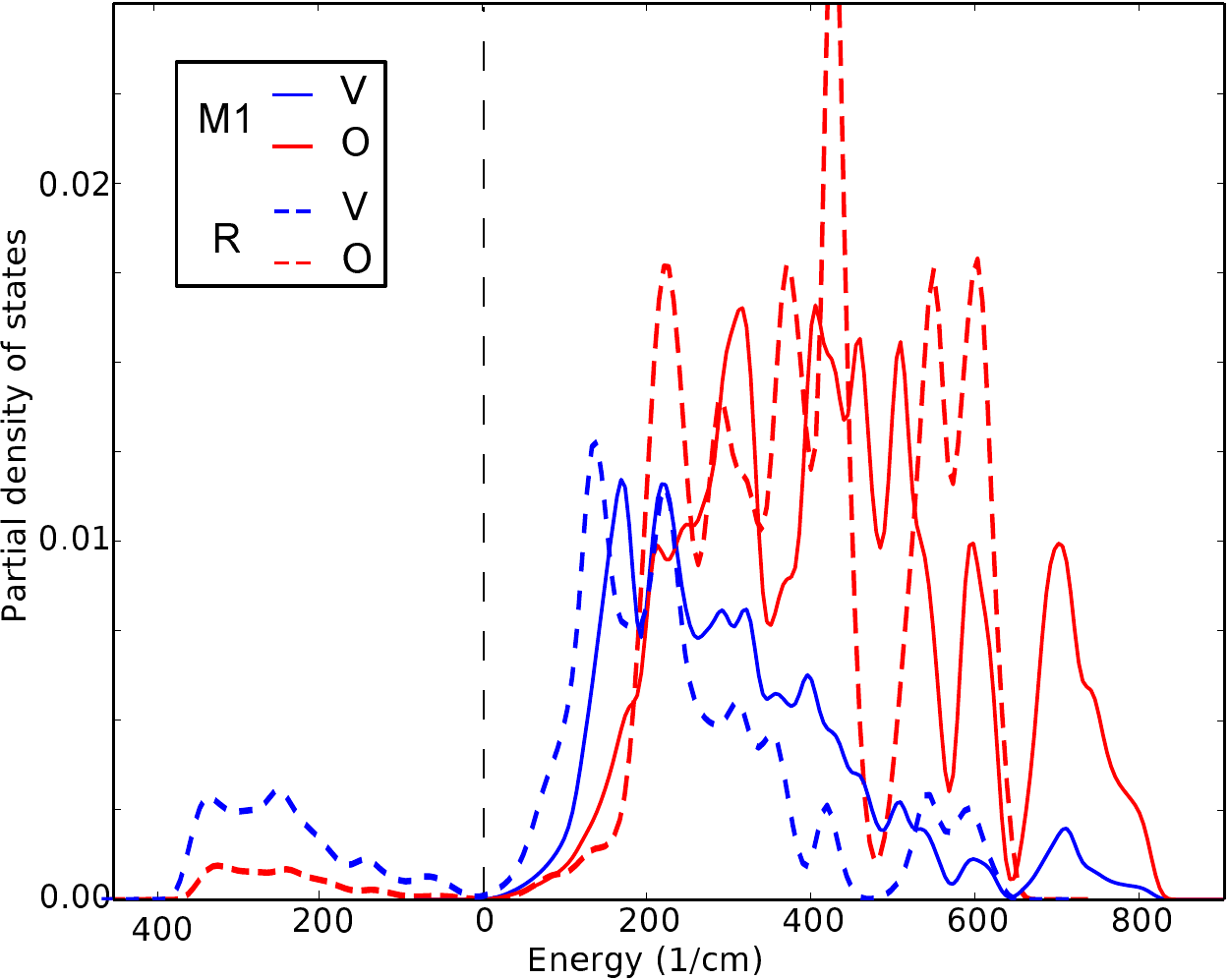}
\par\end{centering}
\caption{{DFT harmonic phonon density of states for M1 and R-VO\protect\textsubscript{2}.
\label{fig:M1 R phonon DOS}}}
\end{figure}

Harmonic DFT phonon dispersion is shown in Fig. \ref{fig:M1 R phonon}.
The Brillouin zones for M1 and R unit cells are sampled between high
symmetry points in reciprocal space. The path for R-VO\textsubscript{2}
follows the sequence $\{Z,\,\Gamma,\,M,\,A,\,Z,\,R,\,X,\,\Gamma\}$
which corresponds to $\{00\frac{1}{2},\,000,\,\frac{1}{2}\frac{1}{2}0,\,\frac{1}{2}\frac{1}{2}\frac{1}{2},\,00\frac{1}{2},\,\frac{1}{2}0\frac{1}{2},\,\frac{1}{2}00,\,000\}$.
The path for M1-VO\textsubscript{2} is $\{\Gamma,\,Y,\,C,\,Z,\,\Gamma,\,B,\,D,\,Z,\,\Gamma,\,A,\,E\}$
which corresponds to $\{000,\,0\frac{1}{2}0,\,\frac{1}{2}\frac{1}{2}0,\,\frac{1}{2}00,\,000,\,0\frac{1}{4}\frac{1}{2},\,\frac{1}{2}\frac{1}{4}\frac{1}{2},\,\frac{1}{2}00,\,000,\,0\frac{1}{4}\bar{\frac{1}{2}},\,\frac{1}{2}\frac{1}{4}\bar{\frac{1}{2}}\}$.
The phonon densities of states for the M1 and R phases are shown projected
by atomic species in Fig. \ref{fig:M1 R phonon DOS}. The eigenvectors
of the imaginary transition modes are shown to project primarily onto
the motion of vanadium atoms.

\section{Thermodynamics\label{sec:thermodynamics}}

The phonon entropy difference between the M1 and the R phases is estimated from the harmonic free energy

\[
S^{\text{ph}}=-\partial_{T}F^{\text{ph}}\,,
\]
where $F^{\text{ph}}$ is 

\[
F^{\text{ph}}=-T\,\text{ln}\,Z\,,
\]
and partition function is computed using the harmonic geometric series expression

\[
Z=\prod_{i \mathbf{q}} \frac{e^{-\beta \omega_{i\mathbf{q}}/2}}{1-e^{-\beta \omega_{i\mathbf{q}}}}\,,
\]
with $\beta\equiv T^{-1}$.


For the M1 phase, entropy is calculated from the standard DFT  harmonic frequencies, $\omega_{i\mathbf{q}}$. For the R phase the same expression is applied to the $3n-2$ real harmonic DFT frequencies that don't soften at the transition, and the two shifted frequencies $\tilde{\omega}_{i\mathbf{q}}$, for the two imaginary harmonic modes subject to the experimental renormalization to real effective frequencies.

The R phase is metallic. As we are only interested in thermal electron
excitations at moderate temperatures we assume
$\partial_{T}g(E)=0$, and that electronic entropy of the R phase
can be given in terms of partial one-electron occupancies as
\[
S^{\text{el}}=\int dE\,g(E)\left\{ f\text{ln}f+(1-f)\,\text{ln}(1-f)\right\} \,.
\]

The total entropy of M1-VO\textsubscript{2} is $S_{\text{M1}}=S_{\text{M1}}^{\text{ph}}$, and the total entropy for R-VO\textsubscript{2} is  $S_{\text{R}}=S_{\text{R}}^{\text{el}}+S_{\text{R}}^{\text{ph}}+\tilde{S}_{\text{R}}^{\text{ph}}$.
$S_{\text{R}}^{\text{ph}}$ is the vibration contribution
from the $3n-2$ phonon modes that are harmonic at low temperature. $\tilde{S}_{\text{R}}^{\text{ph}}$
is from the dispersion two bands that soften at the transition and is based on
the frequencies transformed to $T_{\text{C}}$.

\section{Gaussian Process Regression\label{sec: GPR}}

\begin{figure*}
\begin{centering}
\includegraphics[scale=0.55]{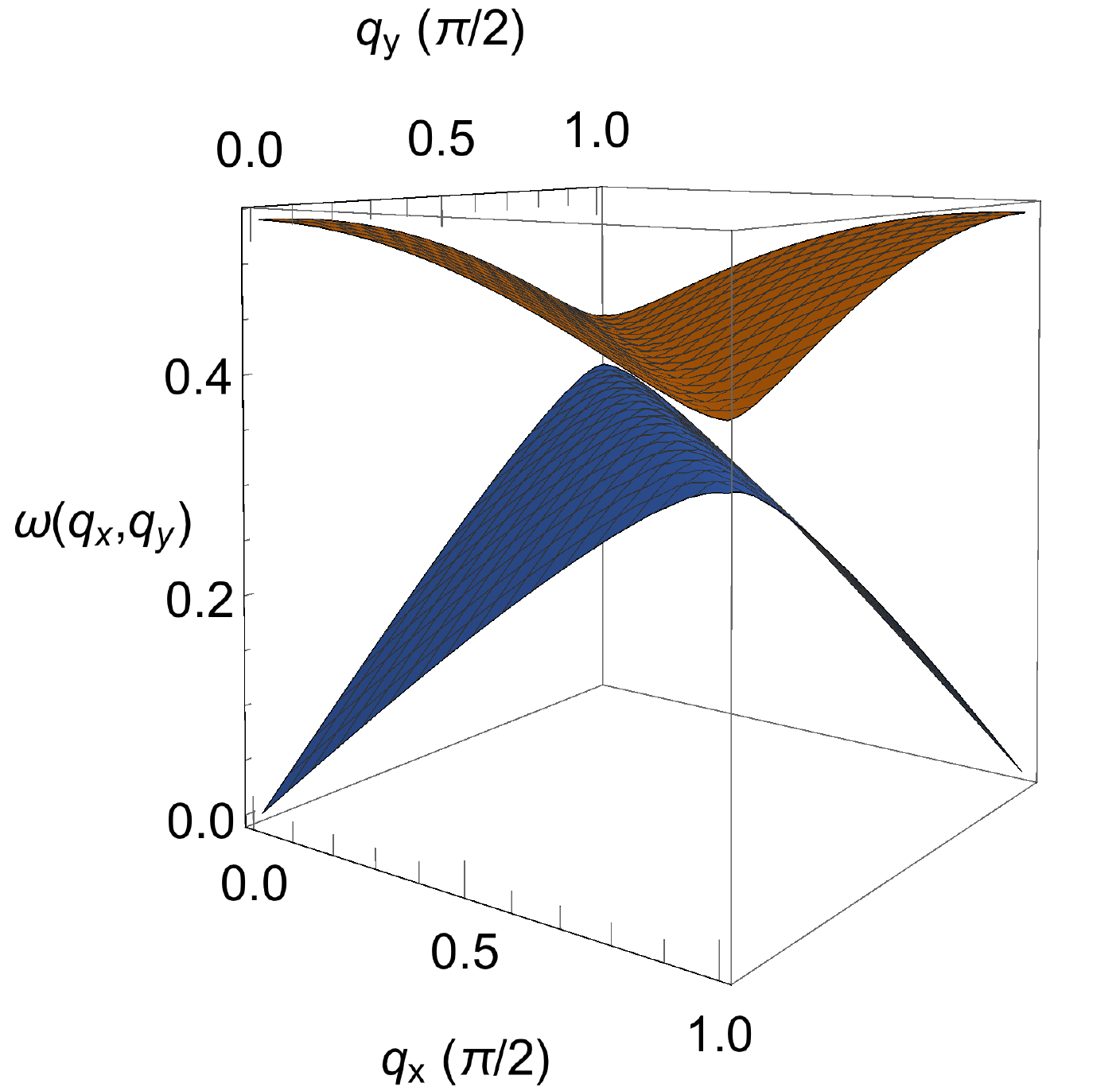}\includegraphics[scale=0.66]{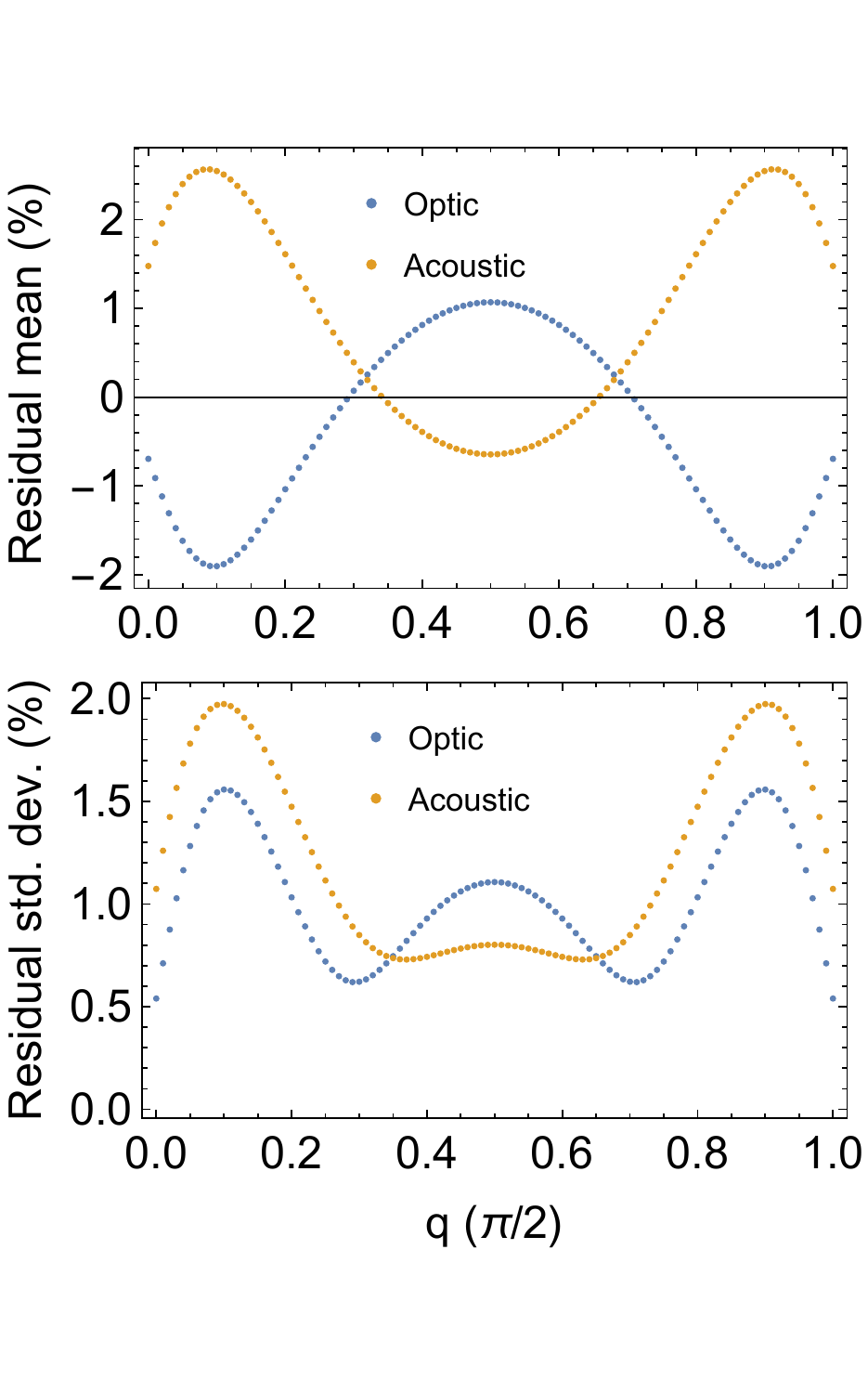}
\par\end{centering}
\caption{{\emph{Right:} Analytic dispersion system used to benchmark 
Gaussian Process Regression (GPR) performance for Brillouin zone interpolation from a limited set of initial data points. \emph{Left:} Mean residual deviation error statistics
for the GPR model in the analytic test system. \label{fig:GPR_analyticModel}}}
\end{figure*}

Supervised learning has been used to interpolate phonon frequencies
using the non-parametric multi-variate Bayesian method Gaussian Process
Regression (GPR).\cite{rasmussen2006gaussian} GPR models can provide an appropriate
alternative to Fourier interpolation, which is otherwise the method
of choice when the full dynamical matrix is known. Without knowledge of the dynamical matrix at the transition, GPR models can
be used to directly interpolate frequencies in \textbf{q} space from
limited $\tilde{\omega}(\mathbf{q})$ experimental data points.

In this work have used GPR for the $\mathbf{q}$-space interpolation
of the two renormalized soft modes in R-VO\textsubscript{2}. The
mode frequencies at $T_{\text{C}}$ are determined from experiment
at limited high-symmetry wavevectors. GPR can be used to predict how
$\tilde{\omega}(\mathbf{q})$ varies across the full Brillouin zone, making
possible thermodynamic calculations for the high-temperature phase
from limited high-temperature data points. To show that the GPR approach
is appropriate to predict the full $\tilde{\omega}(\mathbf{q})$ surface from
limited data points, we benchmark the accuracy of GPR interpolation
on an analytic model.

Consider a vanadium-oxygen analytic model with the following dispersion
relation

\begin{widetext}

\[
\tilde{\omega}(q_{x},\,q_{y})=\sqrt{\left(\frac{1}{m_{\text{O}}}+\frac{1}{m_{\text{V}}}\right)\pm\left\{ \left(\frac{1}{m_{\text{O}}}+\frac{1}{m_{\text{V}}}\right)^{2}-\frac{4}{m_{\text{O}}m_{\text{V}}}\,\text{sin}^{2}\,\mathbf{q}\right\} }\,,
\]

\end{widetext}which is shown in Fig. \ref{fig:GPR_analyticModel}.
The test system includes features such as optic and acoustic-type
dispersion, with frequencies that are non-linear in wavevector in
more than one dimension and that have stationary points of inflection.
The system is therefore expected to provide meaningful accuracy benchmarks,
while also being simple enough to clearly illustrate the method.

In the test system $\tilde{\omega}(q_{x},\,q_{y})$ is sampled by a $100\times100$
mesh over $[0,\,\frac{\pi}{2}]$. GPR training data is a 1D scan of
the mesh of $\tilde{\omega}(q_{x},\,q_{y})$ at the line-paths at $q_{x}=0$
and at $q_{x}=\frac{\pi}{2}$. Root mean square (RMS) residual errors of the interpolated
system compared to the true system are $5$\% for the acoustic band
and $8$\% for the optic, with percentages calculated with respect to
the maximum frequency value of $\tilde{\omega}=0.54$ at $\mathbf{q}=0$.
Typically we also know frequency gradients at zone boundaries. For
a more realistic test model, derivatives at boundaries are included
in the training set. This lowers RMS residual errors across $\mathbf{q}$
to $2$\% and $3$\% for the acoustic and optic bands respectively.

In the GPR applied in this work, for the soft modes in R-VO\textsubscript{2},
an analogous interpolation is made for the two transition bands in
$\{q_{x},\,q_{y},\,q_{z}\}$. Errors of $2$\% for the interpolated
R-VO\textsubscript{2} soft modes correspond to errors of approximately
$1$\% or 0.01 \emph{k}\textsubscript{B}/VO\textsubscript{2} in
the transition entropy difference, which is satisfactory within the scope of this work and in context of other sources of error.

To interpolate the R-VO\textsubscript{2} soft modes we have used a GPR with a non-deterministic
radial basis function kernel of the form

\[
k(q,\,q')=\sigma_{f}^{2}\,\text{exp}\,\frac{-\left(q-q'\right)^{2}}{2l^{2}}+\sigma_{n}^{2}\delta(q,\,q')\,,
\]
with Bayesian maximum posterior $\theta=\{\sigma_{f},\,\sigma_{n},\,l\}$
hyper-parameters. Training data includes line-paths between high-symmetry
points in the $q_{z}=\frac{1}{2}$ and $q_{z}=0$ planes, as well
as zone boundary band velocities. GPR training data for soft modes
in the $q_{z}=\frac{1}{2}$ plane consists of renormalized harmonic
frequencies. These are sampled at 100-points/line for each edge in
the cycle $\{\mathbf{R},\mathbf{Z},\,\mathbf{A},\,\mathbf{R}\}$.
For the $q_{z}=0$ plane, in which the transition-mode bands do not
soften at the transition, training data consists of 100-points/line
samples of the edges in the $\{\mathbf{\Gamma},\mathbf{X},\,\mathbf{M},\,\mathbf{\Gamma}\}$
graph for harmonic frequencies.%

\end{document}